\documentclass[conference]{IEEEtran}
\IEEEoverridecommandlockouts
\usepackage{cite}
\usepackage{amsmath,amssymb,amsfonts}
\usepackage{algorithmic}
\usepackage{graphicx}
\usepackage{textcomp}
\usepackage{xcolor}
\usepackage{comment}
\usepackage[hyphens]{url}
\usepackage[hidelinks]{hyperref}
\def\BibTeX{{\rm B\kern-.05em{\sc i\kern-.025em b}\kern-.08em
    T\kern-.1667em\lower.7ex\hbox{E}\kern-.125emX}}
\begin{document}

\title{Cross-Modal Phantom: Coordinated Camera–LiDAR Spoofing Against Multi-Sensor Fusion in Autonomous Vehicles}

\author{\IEEEauthorblockN{Shahriar Rahman Khan}
\IEEEauthorblockA{\textit{Dept. of Computer Science} \\
\textit{Kent State University}\\
Kent, OH 44242, USA \\
srahmank@kent.edu
\vspace{-5mm}
} 

\and
\IEEEauthorblockN{Raiful Hasan}
\IEEEauthorblockA{\textit{Dept. of Computer Science} \\
\textit{Kent State University}\\
Kent, OH 44242, USA \\
rhasan7@kent.edu
\vspace{-5mm}
}
}

\maketitle

\begin{abstract}
Autonomous Vehicles (AVs) increasingly depend on Multi-Sensor Fusion (MSF) to combine complementary modalities such as cameras and LiDAR for robust perception. While this redundancy is intended to safeguard against single-sensor failures, the fusion process itself introduces a subtle and underexplored vulnerability. In this work, we investigate whether an attacker can bypass MSF’s redundancy by fabricating \textit{cross-sensor consistency}, making multiple sensors agree on the same false object. We design a coordinated, \textit{data-level (early-fusion)} attack that emulates the outcome of two synchronized physical spoofing sources: an infrared (IR) projection that induces a false camera detection and a LiDAR signal injection that produces a matching 3D point cluster. Rather than implementing the physical attack hardware, we simulate its \textit{sensor-level outcomes} by inserting perspective-aware image patches and synthetic LiDAR point clusters aligned in 3D space. This approach preserves the perceptual effects that real IR and IEMI-based spoofing would create at the sensor output. Using 400 KITTI scenes, our large-scale evaluation shows that the coordinated spoofing deceives a state-of-the-art perception model with an 85.5\% successful attack rate. These findings provide the first quantitative evidence that malicious cross-modal consistency can compromise MSF-based perception, revealing a critical vulnerability in the core data-fusion logic of modern autonomous vehicle systems.
\end{abstract}


\section{Introduction}
\label{sec:intro}

Autonomous Vehicles (AVs) are rapidly transitioning from experimental prototypes to commercial use. With this transition, their range of applications continues to expand as they progress from Level-0 toward Level-5 automation. Companies such as Waymo and Tesla are at the forefront of this evolution, Waymo has expanded its driverless taxi services in partnership with Uber \cite{caranddriver2024}, while Tesla has unveiled its fully autonomous Robotaxi platform \cite{reuters2024}. This expansion extends beyond ride-sharing; self-driving buses \cite{spectrum2025}, trucks \cite{ttnews2025}, and delivery vehicles \cite{nbcbayarea2024} are also being actively deployed. However, as AVs become more widespread, incidents involving them have also increased. As of May 2024, the National Highway Traffic Safety Administration (NHTSA) reported 617 AV-related crashes since July 2021, underscoring persistent safety and security challenges\footnote{\url{https://www.nhtsa.gov/laws-regulations/standing-general-order-crash-reporting}}.


\begin{figure}[hbt!]
    \centering
    \includegraphics[width=1\linewidth]{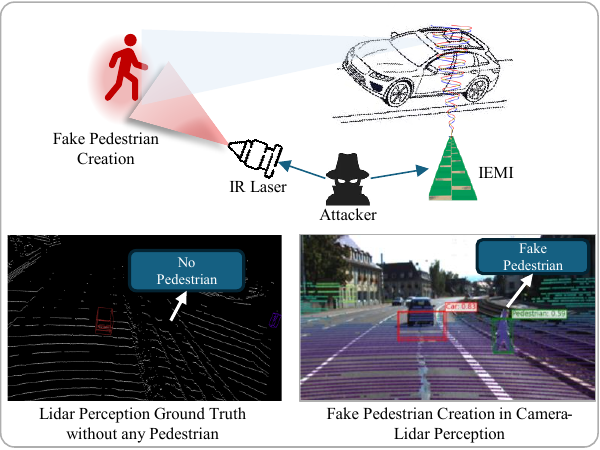}
    \caption{Proposed coordinated, cross-modal attack scenario. A single attacker coordinates two independent spoofing devices: (1) An IR Laser projects an adversarial pattern onto a surface (e.g., road surface) to create a high-confidence 2D camera detection. (2) An IEMI or signal injection device spoofs the LiDAR sensor to create a 3D point cluster. The attack's success depends on synchronizing both signals to correspond to the same 3D spatial location, thus bypassing the MSF's redundancy checks.}
    \label{fig:attack_scene}
\end{figure}

An AV's safety case relies heavily on its perception system, which uses sensors like LiDAR, cameras, and radar to understand the driving environment \cite{li2025multi, zhou2023sensor}. Any attack on this sensor data can directly compromise perception and cascade to catastrophic failures. Consequently, single-sensor perception attacks are a well-studied field, with research demonstrating the feasibility of LiDAR spoofing \cite{bhupathiraju2023emi}, adversarial camera patches \cite{man2020ghostimage}, and radar manipulation \cite{sun2021control}. 

To mitigate the limitations of any single sensor, modern AVs overwhelmingly adopt MSF, combining the strengths of complementary sensors (e.g., camera's rich texture, LiDAR's precise 3D geometry) \cite{biswas2023autonomous}. The core assumption is that fusing data provides robustness; an anomaly in one sensor will be cross-checked and rejected by another. However, this reliance on MSF simultaneously introduces new, complex vulnerabilities. Recent works have shown that adversaries can target the fusion logic itself, for example by creating interference that affects individual sensors to affect MSF perception \cite{cheng2023fusion} or by designing stealthy \textit{Frustum attacks} that maintain semantic consistency across modalities \cite{hallyburton2022security}.

Despite this progress, the literature on AV perception attacks remains fragmented and unknown vulnerabilities persist in MSF systems. This observation motivates a fundamental question: \textbf{is it possible to \emph{bypass} MSF redundancy by fabricating cross-sensor consistency rather than inconsistency?} Can two spoofed modalities \emph{agree on the same illusion}?

Hence, we propose and validate a novel attack vector that does exactly this, creating a high-confidence ``phantom'' object by deceiving both the camera and the LiDAR sensors in a coordinated and spatially synchronized manner, as shown in Figure~\ref{fig:attack_scene}. The main contributions of this paper are as follows:
\begin{itemize}

    \item \textbf{A Novel Threat Model:} We formalize a coordinated, cross-modal attack that fabricates spatial consistency between camera and LiDAR data, exposing a new vulnerability in the fusion logic of MSF systems.
    \item \textbf{Outcome-Equivalent Simulation:} We design a digital simulation pipeline that reproduces the \emph{sensor-level effects} of IR projection and LiDAR spoofing through perspective-aware patch insertion and 3D point injection, enabling safe yet realistic experimentation.
    \item \textbf{Empirical Validation:} Using 400 KITTI scenes, we demonstrate that spatially aligned phantoms deceive the PointPillars perception model with an overall successful attack rate of 85.5\%, showing that cross-modal consistency alone can compromise perception reliability.
    
\end{itemize}

\section{Background and Related Work}
\label{sec:background}


\subsection{Single-Sensor Vulnerabilities}
Individual sensors used by AVs are known to have exploitable vulnerabilities. These attacks form the primitives for our coordinated attack.

\noindent\textbf{Camera-only Perception Attacks:} Camera perception systems typically employ Deep Neural Networks (DNNs) like YOLO \cite{bochkovskiy2020yolov4} or Faster R-CNN \cite{ren2016faster} to process 2D images and generate object detections characterized by bounding boxes, class labels, and confidence scores. However, these optical models are vulnerable to physical-world adversarial examples. Attacks range from applying adversarial patches or stickers to traffic signs to cause misclassification \cite{jia2022fooling, wang2024revisiting}, to projecting light patterns via infrared (IR) lasers \cite{sato2024invisible, man2020ghostimage}. These projections can induce \textit{phantom} detections of non-existent objects by creating adversarial patterns that are invisible to the human eye but salient to the camera sensor.


\noindent\textbf{LiDAR-only Perception Attacks:} LiDAR sensors emit laser pulses to generate sparse 3D point clouds that represent the spatial environment. Perception algorithms such as PointRCNN \cite{shi2019pointrcnn} process this volumetric data to produce 3D object bounding boxes. Despite their precision, LiDAR sensors are susceptible to signal spoofing and point cloud injection. By overpowering the sensor's receiver with a malicious laser, an attacker can inject crafted fake points. This capability allows adversaries to create \textit{phantom} objects \cite{jin2024phantomlidar, jin2023pla, bhupathiraju2023emi}, remove real objects from perception \cite{cao2023you}, or distort the geometric shape of existing obstacles.



\subsection{The Evolution to Multi-Sensor Fusion (MSF) Attacks}
To defend against these single-point failures, modern AVs (from level 3) employ MSF \cite{shen2020drift}. The core assumption is that an anomaly in one sensor (e.g., a ``phantom" LiDAR object) will be cross-checked against the camera feed and, finding no corresponding visual object, will be rejected as noise \cite{biswas2023autonomous}.

However, this fusion logic itself has become the new attack surface. Prior research into MSF attacks has largely focused on creating \textit{inconsistencies} between sensors to confuse the system. For example, some attacks target a single sensor modality with the express purpose of disrupting the fusion output (a ``single-sensor-multi-modal" attack) \cite{cheng2023fusion}. Other works, such as the \textit{Frustum attack}, manipulate data to be semantically consistent (e.g., a valid object type) but physically implausible, aiming to confuse the fusion logic \cite{hallyburton2022security}. A small number of studies have also explored using single physical objects that simultaneously interfere with multiple sensors, such as a 3D-printed object designed to fool both camera and LiDAR \cite{zhu2024malicious, cao2021invisible}.

\textbf{Our Work in Context:} Our proposed attack vector represents a significant departure from this state-of-the-art. Instead of creating \textit{inconsistency} to break the fusion logic, our goal is to create \textit{false consistency} to bypass it entirely. We do not try to make the sensors disagree; we make them \textit{agree} on a non-existent reality. To our knowledge, this work is the first to formally propose and validate a \textit{coordinated, dual-sensor} attack where two \textit{independent, non-physical} spoofing sources (an IR laser for the camera, and LiDAR injection for the LiDAR) are synchronized to create a high-confidence, spatially-consistent phantom object. Our experiments emulate the \emph{sensor-level outcomes} of IR projection and LiDAR return injection (i.e., camera bounding boxes and LiDAR point clusters), rather than reproducing the complete optical or electromagnetic attack hardware. This outcome-emulation approach isolates the vulnerability of fusion and perception logic to plausible sensor outputs; physical reproduction of those outputs is left as future work.

\section{Attack Vector Proposal: Coordinated Cross-Modal Spoofing}
\label{sec:threat}
We propose a novel attack vector that aims to create a consistent, high-confidence \textit{phantom} object detection within an MSF system by simultaneously deceiving both the camera and LiDAR sensors in a spatially synchronized manner. Figure~\ref{fig:attack_scene} illustrates the theoretical concept.

\subsection{Attacker Capabilities and Assumptions}
\label{sec:threat_model}
To formalize the threat, we define the attacker's capabilities:
\begin{itemize}
    \item \textbf{Attacker Goal:} To cause the victim AV to perceive a high-confidence, non-existent \textit{phantom} object at a chosen 3D location ($L_{\text{target}}$), leading to a hazardous action (e.g., sudden braking).
    
    \item \textbf{Attacker Knowledge:} The attacker operates in a \emph{black-box} manner with respect to the internal parameters of the perception and fusion models. However, they are assumed to possess:
    \begin{itemize}
        \item General knowledge of AV sensing hardware (e.g., CMOS cameras, 905\,nm LiDARs) and the fact that MSF systems rely on cross-modal geometric consistency.
        \item The ability to \emph{estimate} the LiDAR--camera extrinsic calibration matrix $T_{ext}$, which is typically not directly exposed to external parties. Prior work shows that such parameters can be inferred through passive observation~\cite{zhang2022play} or model fingerprinting~\cite{sato2024invisible}. 
    \end{itemize}
    
    \item \textbf{Attacker Access:} The attacker is physically present near the victim AV (e.g., on an overpass or roadside) with a line-of-sight to the AV's sensors. They possess two key pieces of hardware:
    \begin{enumerate}
        \item \textbf{IR Laser Projector:} A steerable infrared projector, as demonstrated in \cite{sato2024invisible, man2020ghostimage}, capable of projecting an adversarial pattern onto the AV's camera sensor.
        \item \textbf{LiDAR/IEMI Spoofer:} A device capable of injecting fake point cloud data into the AV's LiDAR sensor, either via direct laser spoofing \cite{jin2023pla} or through electromagnetic interference (IEMI) \cite{bhupathiraju2023emi}.
    \end{enumerate}
    
    \item \textbf{Synchronization \& Physical Constraints:} 
    To maintain cross-modal consistency, the attacker must synchronize the spoofed camera and LiDAR signals within the sensor fusion timing window. 
    In typical AV platforms, camera frames ($10-30$\,Hz) and LiDAR scans ($10-20$\,Hz) are aligned within tens of milliseconds; for example, the Velodyne HDL-64E used in the KITTI dataset operates at $10$\,Hz, producing one full $360^\circ$ scan every $\sim 100$\,ms~\cite{geiger2013vision}. 
    Thus, the attacker must ensure that the spoofed 2D and 3D signals fall within the same fusion cycle. 
    In physical-world execution, this synchronization is further constrained by ego-motion, sensor noise, and environmental factors (e.g., IR attenuation under strong sunlight), requiring the attacker to continuously update projection targets in real time to preserve geometric overlap.
\end{itemize}

\subsection{System and Target Architecture}
We focus on early- and feature-level MSF systems, which represent the state-of-the-art in robust AV perception. Unlike late-fusion systems that compare independent bounding boxes, early-fusion architectures integrate multi-modal data at the feature level before generating final object proposals. Based on the foundational point-level fusion paradigms~\cite{sindagi2019mvx, vora2020pointpainting}, we formalize the target perception pipeline as follows:
\begin{enumerate}
    \item \textbf{Data Acquisition \& Feature Extraction:} The system captures a 2D camera image $I$ and a 3D LiDAR point cloud $PC$. The 2D image is processed to extract visual features (e.g., semantic scores) for each pixel.
    \item \textbf{Spatial Projection (Alignment):} To achieve cross-modal consistency, the algorithm iterates through the 3D LiDAR point cloud. It mathematically projects each 3D point $(x,y,z)$ onto the 2D image plane to find its corresponding 2D pixel coordinate $(u,v)$. This mapping relies entirely on the LiDAR-to-camera extrinsic and intrinsic calibration matrices, collectively denoted as $T_{ext}$.
    \item \textbf{Point-wise Early Fusion:} For each projected 3D point, the algorithm extracts the visual features located at its corresponding $(u,v)$ pixel coordinate. These visual features are directly appended to the 3D LiDAR point, effectively coordinating the geometric point cloud with rich 2D camera data.
    \item \textbf{3D Detection:} This fused, multi-modal point cloud is fed into a 3D object detection network (e.g., PointPillars~\cite{lang2019pointpillars}) to output the final 3D bounding boxes ($D$) and confidence scores.
\end{enumerate}

Our attack targets the core vulnerability of this pipeline: its absolute reliance on the geometric projection ($T_{ext}$) for cross-modal data association. The attacker must generate a set of malicious sensor data ($I_{\text{adv}}$, $PC_{\text{adv}}$) such that, during the \textit{Spatial Projection} phase, the spoofed LiDAR points project exactly onto the pixels containing the spoofed visual features, resulting in a high-confidence detection of a phantom object $O_{\text{phantom}}$ at a chosen 3D location $L_{\text{target}}$.

\subsection{Theoretical Justification of Attack}
We now prove that an attacker can craft the malicious data ($I_{\text{adv}}$, $PC_{\text{adv}}$) to be spatially consistent, thereby fooling the $P_{\text{MSF}}$ model.

\paragraph{Condition 1: Crafting Malicious LiDAR Data ($PC_{\text{adv}}$)}
The attacker chooses a target 3D location and bounding box ($L_{\text{target}}$) for the phantom object.
\begin{itemize}
    \item \textbf{Premise:} As established by prior works \cite{bhupathiraju2023emi, jin2024phantomlidar, jin2023pla}, LiDAR perception models are vulnerable to injected point clouds.
    \item \textbf{Action:} The attacker uses their signal spoofer to generate a set of fake points, $PC_{\text{spoof}}$, that are geometrically consistent with a real object (e.g., a `Pedestrian') at location $L_{\text{target}}$.
    \item \textbf{Result:} The victim's sensor reads $PC_{\text{adv}} = PC_{\text{real}} \cup PC_{\text{spoof}}$. This $PC_{\text{adv}}$ now contains a compelling, realistic 3D representation of an object at $L_{\text{target}}$.
\end{itemize}

\paragraph{Condition 2: Crafting Malicious Camera Data ($I_{\text{adv}}$)}
Simultaneously, the attacker must manipulate the physical environment to make the camera sensor ``see'' the same object.
\begin{itemize}
    \item \textbf{Premise:} As established by prior works \cite{man2020ghostimage, sato2024invisible}, camera perception models are vulnerable to projected light patterns reflecting off physical surfaces.
    \item \textbf{Action:} Using their knowledge of the AV's calibration ($T_{ext}$), the attacker calculates the exact 2D pixel coordinates, denoted $L_{2D} = \text{Proj}(L_{\text{target}})$, that correspond to the fake 3D location. The attacker then aims an IR laser at a physical surface in the real world (e.g., the road surface) to project an adversarial pattern $\delta_I$. The physical projection is calibrated so that the light reflecting into the victim's camera lens forms a 2D image of the phantom object at exactly the target pixels $L_{2D}$.
    \item \textbf{Result:} The victim's sensor captures $I_{\text{adv}} = I_{\text{real}} + \delta_I$. This $I_{\text{adv}}$ now contains a compelling 2D representation of the object perfectly aligned at pixels $L_{2D}$.
\end{itemize}

\paragraph{Condition 3: Bypassing the Fusion Logic (Satisfied)}
The attacker has now created a perfect, data-level illusion synchronized across the physical environment.
\begin{itemize}
    \item \textbf{Proof:} The AV's MSF pipeline receives $I_{\text{adv}}$ and $PC_{\text{adv}}$. During the \textit{Spatial Projection} phase, the fusion algorithm takes the spoofed 3D points at $L_{\text{target}}$ and projects them onto the 2D image plane using $T_{ext}$. Because the attacker mathematically derived the physical IR laser projection to land exactly on pixels $L_{2D}$ using the same geometric relationship ($T_{ext}$), the injected 3D points project onto the spoofed 2D visual patch. Consequently, during the \textit{Point-wise Early Fusion} phase, the algorithm extracts the malicious visual features and actively binds them to the injected 3D points. This creates a multi-modal feature set that is mathematically indistinguishable from a real object, forcing the 3D detection network to output a high-confidence false detection $O_{\text{phantom}}$.
\end{itemize}
\section{Experimental Validation Setup}
\label{sec:setup}
To empirically validate the impact of our proposed physical attack, we conduct a high-fidelity \textit{digital simulation} using the KITTI benchmark dataset \cite{geiger2012we}. Our simulation is designed to test the most critical component of our threat model: the effectiveness of the coordinated, multi-modal spoofing data ($I_{\text{adv}}$ and $PC_{\text{adv}}$) against a state-of-the-art early-fusion perception pipeline.

\subsection{Justification of Simulation Methodology}
A full, end-to-end simulation of a physical attack on a complete MSF pipeline ($P_{\text{MSF}}$) is computationally complex, requiring precise physics modeling of both IR laser propagation and IEMI signal injection. Instead, we perform a closed-lab digital simulation experiment to validate the core logic of the attack vector itself. 

We simulate the outcome of the physical attack at the data level. Our script performs both steps of the attack synchronously:
\begin{enumerate}
    \item \textbf{Simulating Condition 1 (Camera):} We generate the malicious camera image $I_{\text{adv}}$ by programmatically inserting a 2D object patch at the projected location $L_{2D}$. This models the data-level outcome of a successful IR laser attack.
    \item \textbf{Simulating Condition 2 (LiDAR):} We create the malicious point cloud $PC_{\text{adv}}$ by injecting a corresponding 3D point cluster into the raw `.bin' file.
\end{enumerate}

We then project the 3D LiDAR data onto the 2D camera image using the calibration matrices to align the modalities. This combined, malicious data is fed into a state-of-the-art perception model relying on PointPillars \cite{lang2019pointpillars}. If this model is successfully fooled (i.e., it outputs a high-confidence phantom detection), we confirm the fusion logic is bypassed. We acknowledge that this image-patch experimental approach is a simplification; it assumes a successful projection without modeling complex environmental factors (e.g., ambient light, surface reflectivity). Hence, our future work will focus on employing physics-based environmental rendering to rigorously simulate these physical phenomena.

\subsection{Implementation Details}
\label{sec:implementation}
\textbf{Dataset:} We use the KITTI 3D Object Detection dataset \cite{geiger2012we}. We built a library of source objects (`Car' and `Pedestrian') from various training scenes to ensure realistic phantom morphologies.

\textbf{Augmentation Pipeline:} We developed a Python script that precisely implements our digital attack. We conducted a large-scale experiment, creating $400$ augmented scenes. This included $200$ scenes injected with a `Pedestrian' phantom and $200$ scenes with a `Car' phantom, each placed at a random, plausible location ($20-40$m in front of the AV).

\textbf{Perception Model ($P_{\text{MSF}}$):} We use a pre-trained early-fusion pipeline utilizing PointPillars \cite{lang2019pointpillars} as our target perception module. This represents a standard, high-performance baseline used in AV perception research.

\textbf{Evaluation Metric:} To evaluate the attack, we define the Attack Success Rate (ASR). An attack is considered ``successful'' if the target model processes the fused adversarial data and outputs a detection of the correct class (matching the injected phantom) with a confidence score $> 0.5$, and the predicted bounding box overlaps with the attacker's intended 3D location.
\section{Experimental Results and Analysis}
\label{sec:results}
To validate our multi-modal augmentation pipeline, we conducted a large-scale experiment by injecting objects from the two primary KITTI classes: \textit{Car} and \textit{Pedestrian}, into $200$ unique scenes for each class. The resulting augmented data was then processed by a pre-trained PointPillars model to evaluate Detection Rate (DR), which is the percentage of augmented objects successfully detected by the model.

\subsection{Overall Attack Success Rate}
\begin{figure}
    \centering
    \includegraphics[width=1\linewidth]{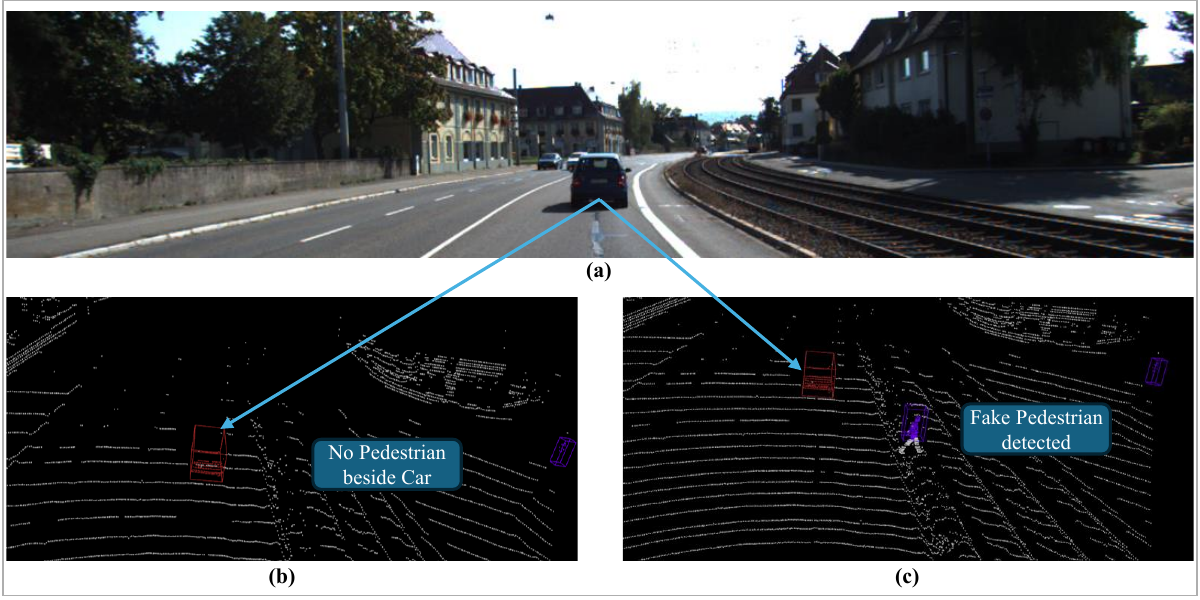}
    \caption{Validation of the LiDAR Spoofing Attack (Condition 2). This figure contrasts the perception output before and after the injection of a phantom point cloud. (a) The original, unmodified camera view of the target scene. (b) The LiDAR point cloud map of the original scene. The perception model correctly detects the real \textit{Vehicle} (red box) and confirms there is `No \textit{Pedestrian} beside \textit{Vehicle}'. (c) The LiDAR point cloud map of the augmented scene. After injecting the phantom point cloud, the perception model now detects both the real \textit{Vehicle} and the `\textit{Phantom Pedestrian}' (purple box), confirming the successful injection of a high-confidence phantom object into the LiDAR sensor stream.}
    \label{fig:result2}
\end{figure}

The coordinated attack proved to be highly effective, achieving an $85.5\%$ overall successful attack rate across all $400$ attempts ($342$ successful detections). The model consistently detected the coordinated phantom objects as real, high-confidence objects. This successful injection of a phantom object into the LiDAR sensor stream is visualized in Figure~\ref{fig:result2}(c), whereas Figure~\ref{fig:result2}(a) and (b) shows the camera view and LiDAR view of the scene, respectively. 

As detailed in Table~\ref{tab:results}, the attack success rate was high for both classes. The \textit{Vehicle} phantoms, which are large and have dense point clouds, were successfully detected in $176$ of $200$ attempts, achieving an $88.0\%$ Detection Rate (DR). `Pedestrian' phantoms, which are smaller and sparser, were also highly successful, being detected in $166$ out of $200$ attempts for an $83.0\%$ DR. This small performance gap is expected, as \textbf{Vehicle} objects provide a stronger, more robust signal for the detector to lock onto. 

\begin{table}[hbt!]
\renewcommand{\arraystretch}{1.3}
\centering
\caption{Attack Success Rate and Confidence Scores by Injected Object Class}
\label{tab:results}
\resizebox{\columnwidth}{!}{
\begin{tabular}{lccccc}
\hline
\textbf{Object Class} & \textbf{Attack} & \textbf{Successful} & \textbf{Attack Success} & \textbf{Mean} & \textbf{Median} \\
& \textbf{Attempts} & \textbf{Phantom Detections} & \textbf{Rate (\%)} & \textbf{Score} & \textbf{Score} \\ \hline
Vehicle & 200 & 176 & 88.0\% & 0.75 & 0.79 \\
Pedestrian & 200 & 166 & 83.0\% & 0.56 & 0.56 \\ \hline
\textbf{Total (Observed)} & \textbf{400} & \textbf{342} & \textbf{85.5\%} & \textbf{0.66*} & \textbf{N/A} \\ \hline
\end{tabular}
}
\par
\textit{*Weighted average of successful attack detection scores.}
\end{table}
 
\subsection{Qualitative Analysis of a Successful Attack}
\begin{figure}[hbt!]
    \centering
    \includegraphics[width=1\linewidth]{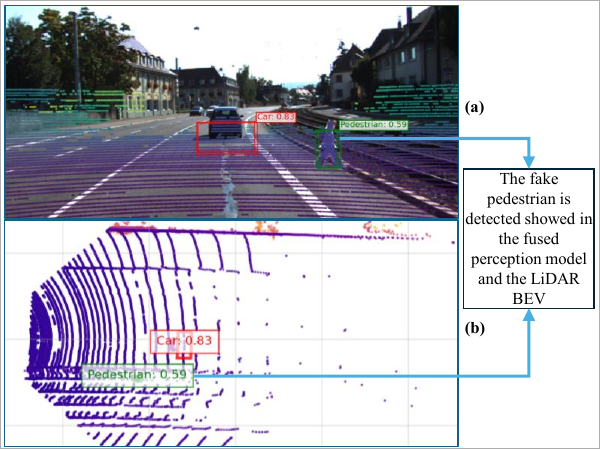}
    \caption{A representative result of a successful attack. (a) The camera-LiDAR fusion perception view, where a synthetic pedestrian patch (simulating the IR laser) and the LiDAR point cloud of the fake pedestrian (through IEMI signal) has been injected. (b) The corresponding LiDAR Bird's-Eye-View (BEV), where the spatially-aligned phantom point cloud was injected. The fusion model is successfully deceived: it correctly identifies the real `Car' (score 0.83) and, crucially, also detects the non-existent \textit{Phantom Pedestrian} (score 0.59) as a valid object.}
    \label{fig:result}
\end{figure}

Figure~\ref{fig:result} provides a representative qualitative example of a successful attack on a particular scene from KITTI dataset. Figure~\ref{fig:result}(a) shows the camera-LiDAR fusion perception view, where a synthetic pedestrian patch (simulating the IR laser) and the LiDAR point cloud of the fake pedestrian (through the IEMI signal) have been injected. The Figure~\ref{fig:result}(b) image shows the corresponding LiDAR Bird's-Eye-View (BEV), where the synthetic point cluster has been injected.

The PointPillars model processes this fused malicious data and produces two detections: correctly identifies the real \textit{Car} (score $0.83$) and, more importantly, \textbf{incorrectly identifies the phantom pedestrian} (score $0.59$). This single example demonstrates the attack's core success: the spatially-aligned phantom data are accepted by the fusion model as a legitimate object with high confidence.
 
\subsection{Analysis of Detection Confidence}
A simple success/fail metric is insufficient; a low-confidence detection might be filtered out by the AV's planning module. Our analysis shows that this is not the case. The phantom objects were detected with alarmingly high confidence.

\textbf{For \textit{Vehicle} phantoms,} the mean confidence score for all $176$ successful attack detections was \textbf{0.75}, with a median of \textbf{0.79}, as shown in Table~\ref{tab:results}. This is an exceptionally high score, well above any reasonable perception threshold. In fact, the confidence distribution was strongly skewed toward high values, with the vast majority of detections clustering between $0.7$ and $0.9$.

\textbf{For \textit{Pedestrian} phantoms,} the mean confidence was \textbf{0.56}, with a median of \textbf{0.56}, as shown in Table~\ref{tab:results}. While lower than \textit{Vehicle} again, likely due to the sparser geometry, this score is still significantly high and would be considered a valid, high-priority object by any downstream module in AV pipeline. This was not an anomaly; the confidence distribution showed a strong peak around $0.7$, confirming that these were not low-quality outliers. These high scores prove that the PointPillars model is confidently fooled, demonstrating the robustness of our proposed coordinated attack on MSF perception.
\section{Conclusion}
\label{sec:conclusion}
This paper identified a critical and under-explored vulnerability in MSF systems utilized in modern AV perception. We moved beyond single-sensor attacks to propose and analyze a coordinated, cross-modal attack that targets the fusion logic itself. We analytically demonstrate that by synchronizing a \textit{phantom} camera detection (via IR laser) with a \textit{phantom} LiDAR detection (via spoofing), an attacker can satisfy all conditions to create a high-confidence, fused, and false object detection. Our large-scale empirical validation, running $400$ simulated attacks, confirms this threat. We demonstrate that a state-of-the-art PointPillars model is successfully deceived with an overall successful attack rate of $85.5\%$. Crucially, these false detections were registered with alarmingly high confidence: averaging $0.75$ for vehicles and $0.56$ for pedestrians, scores that are well above typical perception thresholds.

However, we explicitly acknowledge a primary limitation of this work: the validation relies highly on digital simulation. While this experiment successfully proves the vulnerability in the fusion logic, it cannot fully capture the stochastic complexities of the physical world such as sensor noise, precise hardware and timing synchronization, atmospheric interference and dynamic lighting conditions. Consequently, the critical next step for this line of research is to bridge the gap between simulation and reality by focusing on constructing the physical attack hardware, specifically integrating a synchronized IR laser projector and LiDAR signal injector, to demonstrate the attack's feasibility and robustness in real-world driving settings. Furthermore, this work serves as a call to action for the AV safety community to develop fusion algorithms that are resilient to such coordinated consistency attacks.

\bibliographystyle{IEEEtran}
\bibliography{8_references}

\end{document}